# Unusual dependence on the angle of magnetic field for the spin Hall magnetoresistance of monodomain epitaxial BiFeO$_3$ thin films

Yongjian Tang[1], Pratap Pal[2], Matthew Roddy[1], Jon Schad[2], Ruofan Li[1], Joongwon Lee[3], Farhan Rana[3], Tianxiang Nan[4], Chang-Beom Eom[2], and Daniel C. Ralph[1,5*]

[1]Department of Physics, Cornell University, Ithaca, New York 14853, USA.

[2]Department of Materials Science and Engineering, University of Wisconsin-Madison, Madison, Wisconsin 53706, USA.

[3]School of Electrical and Computer Engineering, Cornell University, Ithaca, New York 14853, USA.

[4]School of Integrated Circuits and Beijing National Research Center for Information Science and Technology (BNRist), Tsinghua University, Beijing 100084, China

[5]Kavli Institute at Cornell, Ithaca, New York 14853, USA

*Corresponding Author: Daniel C. Ralph, Email: <dcr14@cornell.edu>

ABSTRACT: Spin Hall magnetoresistance (SMR) measurements provide a way to probe the surface spin structure of insulating magnetic materials. Such measurements produce resistance signals of the form $\Delta R \propto \cos[2(\alpha-\alpha_0)]$, where $\alpha$ is the angle between the current and the external in-plane magnetic field. Previous experiments on a wide range of materials have found $\alpha_0 = 0°$ for ferromagnets and $\alpha_0 = 90°$ for antiferromagnets. Here we investigate SMR in bilayers of Pt with monodomain $BiFeO_3$ multiferroic epitaxial thin films. We observe signals of the form $\Delta R \propto \cos[2(\alpha-\alpha_0)]$ but surprisingly the angle $\alpha_0$ can take values very different from 90° or 0°, with large variations from sample to sample. The aim of the paper is to report this striking departure from the expected magnetic field dependence of SMR and to encourage consideration of possible microscopic mechanisms.

1. Introduction

Spin Hall magnetoresistance (SMR) [1-10] is a transport phenomenon arising at the interface between an insulating magnetic layer and a metallic nonmagnetic layer with strong spin-orbit coupling, usually a heavy metal (e.g., Pt). Spin current generated by the spin Hall effect in the heavy metal layer reflects from the magnetic interface with an amplitude that can depend on the magnetic orientation, and then is transduced back into a charge current (adding to the originally applied charge current) by the inverse spin Hall effect in the heavy metal. The result is a magnetoresistance signal that depends on the angle of an in-plane magnetic field with the form $\Delta R \propto \cos[2(\alpha-\alpha_0)]$ where $\alpha$ is the angle between the current and the external in-plane magnetic field. For a ferromagnetic layer, $\alpha_0 = 0°$ (also known as positive SMR), because the spin current from the spin Hall effect is absorbed most efficiently when the magnetization is aligned with the charge current. For previous measurements utilizing antiferromagnetic layers $\alpha_0 = 90°$ (known as negative SMR), which is understood as due to a tendency of an applied magnetic field to orient the spin sublattices within an antiferromagnet perpendicular to the field (i.e., in a spin-flop configuration) as schematically illustrated in Fig. 1. This behavior has been observed with a variety of antiferromagets [5-10], including multiferroic $BiFeO_3$ (BFO) samples not containing a single ferroelectric/ferroelastic domain [11-13].

In this work, we investigated the magnetic field dependence of the magnetoresistance in bilayer samples in which a layer of monodomain multiferroic BFO [14-16] formed by deposition on a miscut substrate is integrated with a Pt layer. We observe SMR signals with the signature field dependence $\Delta R \propto \cos[2(\alpha-\alpha_0)]$, but very surprisingly $\alpha_0$ can differ from 0° or 90°, with variation over a large range in different samples, even among different devices made simultaneously on the same wafer. We cannot yet offer a convincing microscopic explanation for this result, but we wish to bring it to the attention of the spintronics community as a striking exception to the established understanding of SMR.

## 2. Growth and Characterization of the BFO Films

For our studies we utilized (001) monodomain BFO thin films grown on SRO-coated cubic SrTiO$_3$ (STO) (001) single-crystal substrates engineered with a 4° miscut toward the [110]$_{pc}$ direction [17-19]. Figure 2 shows characterization results for the BFO films using x-ray scattering and atomic force microscopy. Room-temperature X-ray diffraction (Fig. 2a) shows BFO and SRO peaks around the STO (002) peak. The rocking curve for the (002) peak of BFO (Fig. 2c) has a FWHM of $\Delta\omega$~0.2° which is comparable to good-quality thin films reported previously [20,21]. An epitaxial relationship between the substrate and the BFO films was confirmed by off-axis azimuthal $\phi$-scans around the symmetric (103) reflection for both STO substrate and BFO film as shown in Fig. S1 (see supplementary material). To check for ferroelastic domain variants, we performed reciprocal space mapping (RSM) focusing about the ($\bar{1}$03) peak, as shown in Fig. 2d, which highlights a single peak for the BFO with no trace of minority domains. Similar results were also obtained for the (002) and (113) peaks along both the miscut and non-miscut directions as shown and discussed in Fig. S3 (see supplementary material). Like previous studies on similarly prepared films [17,18], we therefore conclude that our BFO films contain a single ferroelastic domain. Figure S4 of the supplementary material shows optical second harmonic generation measurements which provide additional confirmation for this conclusion. Atomic force microscopy (Fig. 2b) indicates a root-mean-square surface roughness of ~ 3 nm.

To determine the virgin ferroelectric state of the BFO samples we fabricated (ex situ) 30-nm-thick Pt circular electrodes with various diameters diameter (25–200 µm) on top of the BFO and performed measurements of polarization versus electric field (*P-E* loop) of as-deposited samples. Positive electric field is defined as pointing downward (from Pt toward SRO). First, we have measured without any preset electric pulse as shown in Fig. S2 (see supplementary material). The polarization state is unchanged upon sweeping in the positive electric-field direction (corresponding to downward pointing electric field), but then switches when a negative field swept beyond the coercive field (Fig. S2b,c). The magnitude of this switching corresponds to the saturated polarization value measured in the full *P-E* loop (Fig. 2e). From these measurements we conclude that the polarization in the as-deposited state has a uniform downward-pointing out-of-plane component, as also previously demonstrated [17].

A previous neutron-diffraction study [18] on 300 nm BFO films prepared the same way in the same deposition system as our samples determined that the magnetic configuration of the BFO is a single antiferromagnetic domain containing a cycloid with a propagation direction in the film plane parallel to [1$\bar{1}$0] and with a period of 66 ± 2 nm. The cycloid plane lies 12° from the film plane. The polarization vector ***P*** therefore does not lie within the cycloid plane. The magnetic configuration is different from bulk BFO in which a cycloid can exist within any of the three {112}-type planes and the cycloid plane contains ***P*** [22], and different from thinner epitaxially-strained BFO films in which no cycloid is present and the Néel vector is oriented perpendicular to ***P*** [23].

3. Results: Angle-Dependent Magnetoresistance Measurements

To perform the SMR studies, we deposited 10 nm Pt films on virgin BFO films and patterned Hall structure in the Pt/BFO bilayers. The conducting SRO layer beneath the BFO makes no contribution to the SMR measurements because the BFO is highly insulating, giving a resistance between the Pt overlayer and the SRO greater than 32 MΩ.

3.1. Dependence on magnetic-field angle sweeping away from the sample plane

The sample geometry for the angle-dependent magnetoresistance measurements is shown by the diagrams in Fig. 3. We define the x axis as parallel to the current channel in the Hall bar, and z as the out-of-plane direction. The orientation of the spin polarization generated by the spin Hall effect in the Pt is along the y axis. Panels (a) and (b) in Fig. 3 show the longitudinal resistance $R_{xx}$ and the relative change in resistance $(R_{xx} - R_0)/R_0$ (where $R_0$ is the zero-field resistance) measured as a function of magnetic-field angle at fixed magnitudes of magnetic field from 2 T to 8 T for one of our Pt (10 nm)/BiFeO$_3$ (300 nm) samples, with the field angle scanned within the yz and xz planes. For this sample the charge current is applied along the [1$\bar{1}$0] axis of the BFO, i.e., along the direction of the spin-cycloid propagation. For the yz-plane scan, the resistance has the form

$$\frac{\Delta R}{R_0} = SMR_{long} \cos 2\beta,$$

where $\Delta R = R_{xx} - R_0$, $SMR_{long}$ is a prefactor, and the angle $\beta$ is as defined in Fig. 3(a). This relative change reaches a minimum when the external field is normal to the sample plane and a maximum when the external field is along the y axis. The magnetoresistance amplitude in the xz-plane scan is much weaker, about an order of magnitude smaller than that in the yz-plane scan. These dependencies are as expected for a signal dominated by SMR in a HM/AF bilayer in which the effect of the magnetic field is to orient at least a fraction of the spins within the AF layer perpendicular to the field, so that in the yz-plane scan the fraction of spins flopped into the y direction is minimized when the applied field is in-plane (parallel to the y axis). The magnetoresistance for the scan in the xz plane is not strictly zero as would be expected from a signal entirely due to SMR for a well-aligned magnetic field. This could be due to field misalignment from the xz plane or to a small anisotropic magnetoresistance due to induced interfacial magnetism. In magnetization measurements using a SQUID, we observe a small reorientable magnetic moment in the BFO/Pt samples that we associate with interfacial magnetism, a signal that is not present for BFO samples with no Pt layer (see Fig. S5 in supplementary material).

The dependence of the SMR amplitude on the magnitude of applied magnetic field in the yz-plane scan is shown in Fig. 3c. The SMR ratio increases approximately quadratically with field magnitude up to 8 T. Similar behavior is also seen commonly in other HM/AF systems. The longitudinal SMR signal in an antiferromagnet with two spin sublattices is generally modeled as

$$\rho_{xx} = \rho_0 + \frac{\rho_s}{2}\left(2 - m_{1,y}^2 - m_{2,y}^2\right)$$

where $m_{1,y}$ and $m_{2,y}$ are the sublattice magnetization components and $\rho_s$ is the material-specific SMR coefficient [6]. If the local values of $m_{1,y}$ and $m_{2,y}$ depend linearly on the magnetic field magnitude at low field, it is natural that the SMR amplitude is quadratic with field. This behavior has most commonly been observed in samples containing multiple antiferromagnetic domains where the effect of the applied field is to shift domain walls [6,8]. However, this behavior is more general, and will also hold for a sample containing a spin cycloid (see supplementary material). In fact, it should likely hold for any material for which the average value of $m_y$ depends linearly on the magnitude of magnetic field.

At $H = 8\,T$, the SMR amplitude for the device in Fig. 3 is $SMR_{long} = 1.70 \times 10^{-3}$. This is comparable to previous SMR measurements on YIG/Pt [2] and $\alpha$-Fe$_2$O$_3$/Pt [10] and much larger than for NiO/Pt [24]. We interpret the large SMR amplitude as an indication of a high-quality interface between the Pt and BiFeO$_3$.

3.2. Dependence on magnetic-field angle sweeping within the sample plane

The magnetoresistance scans within the xy plane are where our observations differ markedly from previous measurements of SMR in heavy metal/antiferromagnet samples. The external magnetic field rotates in the sample plane with a field angle α measured clockwise from the current (Fig. 4a). Figure 4b shows the relative change in longitudinal resistance measured as a function of magnetic-field angle within the xy plane at fixed magnitudes of magnetic field from 2 T to 8 T, for the same Pt (10 nm)/BiFeO$_3$ (300 nm) sample highlighted in Fig. 3, for which charge current is applied along the [1$\bar{1}$0] axis. Figures 4c-e show similar plots for three other devices on the same sample wafer, for which charge current is applied at the angle $\varphi_{C,1-10}$ = 45°, 75°, 90° away from [1$\bar{1}$0], respectively. In each case, the signals retain the angular dependence ΔR = $A$cos[2(α-α$_0$)] where $A$ is an angle-independent constant, but with values of α$_0$ that differ from the expected value of 90°, and that also vary from device to device: for the $\varphi_{C,1-10}$ = 0° we have α$_0$= 120.3°, for $\varphi_{C,1-10}$ = 45° α$_0$= 19.6°, for $\varphi_{C,1-10}$ = 75° α$_0$ = 172.8°, and for $\varphi_{C,1-10}$ = 90° α$_0$= 66.6°. Each measured angle has an uncertainty of roughly 10° due to sample mounting. However, the differences in α$_0$ cannot be ascribed to errors in mounting or the mechanical rotator as these samples were on the same sample wafer measured together without removal from the rotator. As is visible in Fig. 4, the in-plane SMR amplitudes $A$ also show large sample-to-sample variations, by as much as a factor of 5.

We have performed similar measurements on 13 devices from three different sample wafers. The distribution for the values of α$_0$ for all of the samples and their associated SMR magnitudes are shown in Fig. 5a. All but one sample show large differences from the anticipated angle $\alpha_0$ = 90°, with no apparent underlying pattern. (As noted above, we estimate uncertainties in rotation angles between different sample wafers loaded at different times onto the rotator of at most 10°.) To check whether there is any correlation between $\alpha_0$ and the orientation of the magnetic field relative to the crystalline axes, in Fig. 5b we show the distribution of magnetic-field angles corresponding to maximum SMR measured relative to the [1$\bar{1}$0] crystal axis. Again, there is a large spread in values.

4. Discussion

The main purpose of this paper is simply to point out this unusual behavior and encourage readers to consider what might be the microscopic cause. We have considered several possible models, none of which appear to provide a persuasive explanation.

*Tilting of the cycloid plane ralative to the sample plane.* Strong magnetic anisotropy associated with the cycloid plane in a monodomain BFO sample could cause spin reconfigurations to depend most senstively on the component of magnetic field along the the axis defined by the intersection of the cycloid plane and the sample plane, rather than being uniformly sensitive to any direction of in-plane magnetic field. This would shift the value of $\alpha_0$ away 90°, but for this mechanism the field angle corresponding to the maximum SMR signal should be the same relative to the crystal axes within all monodomain BFO samples. This is not what we observe.

*Domain formation.* Even though our x-ray diffraction and optical second harmonic generation measurements strongly indicate that the bulk of our BFO films consists of a single ferroelectric domain, we have considered whether domain formation of the spin texture could explain our results. Previous work has shown that in antiferromagnetic samples containing multiple domains with spin configurations determined by crystalline anisotropies, $\alpha_0$ remains equal to 90° despite the presence of the domains as long as there are 3 or more domain variants for which the projections of the anisotropy axes onto the sample plane are equally spaced [6]. If there are only two domain variants differing by 90°, or if the projections of the anisotropy axes onto the sample plane are not equally spaced, this picture breaks down and $\alpha_0$ can exhibit values different from 90° (see supplementary material section S2). However, also in this case the field angles corresponding to maximum SMR should be correlated with crystal axes, which is not consistent with our measurements.

*Surface spin texture different than bulk spin texture.* We are aware of only one previous observation of values of $\alpha_0$ different from 0° or 90° – in a narrow range of temperature corresponding to a skyrmion lattice phase in the chiral magnetic insulator $Cu_2OSeO_3$ [25]. The phase changes in that case were ascribed to surface twists in the spin configuration. Something similar could be happening in our samples, in that the presence of a Pt overlayer might cause the surface spin state in our BFO to be different than the monodomain state that exists in the bulk of the BFO film. However, within this scenario it remains unclear why there should be large sample-to-sample differences even in samples from the same wafer.

5. Conclusions

We have investigated the spin Hall magnetoresistnace (SMR) of bilayers consisting of a monodomain multiferroic BiFeO$_3$ (BFO) thin film with a thin Pt overlayer. The SMR in the out-of-plane $β$ scan resembles that of other antiferromagnetic insulators interfaced with a heavy metal. Nevertheless, unlike previous studies of SMR in heavy-metal/ferromagnet and heavy-metal/antiferromagnet samples, the dependence of the SMR on magnetic-field angle is unusual. It goes as $\Delta R \propto \cos[2(\alpha-\alpha_0)]$ where $\alpha$ is the angle of the in-plane magnetic field relative to the applied current, but with values of $\alpha_0$ very different from previous measurements for which $\alpha_0$ takes only the values 0° or 90°. We have considered whether this unusual behavior might be explained by tilting of the cycloid plane in the BFO relative to the sample plane, by domain formation, or by differences in spin configuration between the Pt/BFO interface and the bulk of BFO, but in all of these scenarios the dependence of the SMR on magnetic field angle should presumably be correlated with the crystal axes of the BFO, which is not what we observe. We welcome further consideration by readers about why the magnetic-field dependence of SMR for Pt/(monodomain BFO) samples differs so strikingly from samples made using other insulating antiferromagnets.

**Data statement**

Data are deposited in the Zenodo repository doi.org/10.5281/zenodo.15672539.

**Notes**

The authors declare no competing financial interest.

**Author contributions**

**Yongjian Tang**: Conceptualization, Formal Analysis, Investigation (sample fabrication & transport measurements), Writing – original draft  **Pratap Pal**: Formal Analysis, Investigation (BFO growth and characterization), Writing – original draft  **Matthew Roddy**: Formal Analysis  **Jon Schad**: Investigation (BFO growth and characterization)  **Ruofan Li**: Investigation (sample fabrication)  **Joongwon Lee**: Investigation (optical measurements)  **Farhan Rana**: Supervision  **Tianxiang Nan**: Investigation (sample fabrication)  **Chang-Beom Eom**: Supervision  **Daniel C. Ralph**: Supervision, Writing – original draft  **all authors**: Writing – review and editing.

**Funding Sources**

Y.T. was supported by the U.S. Department of Energy (DE-SC0017671). C.B.E. acknowledges support for this research through a Vannevar Bush Faculty Fellowship (ONR N00014-20-1-2844), and the Gordon and Betty Moore Foundation's EPiQS Initiative, Grant GBMF9065. Ferroelectric measurements at the University of Wisconsin–Madison were supported by the US Department of Energy (DOE), Office of Science, Office of Basic Energy Sciences (BES), under award number DE-FG02-06ER46327. M.R. was supported by SUPREME, one of seven centers in JUMP 2.0, a Semiconductor Research Corporation (SRC) program sponsored by
13

DARPA. R.L. was supported by the Cornell Center for Materials Research (CCMR), part of the National Science Foundation MRSEC program (DMR-1719875). The research at Cornell made use of the CCMR shared facilities and the Cornell NanoScale Facility, a member of the National Nanotechnology Coordinated Infrastructure (NNCI), which is supported by the National Science Foundation (grant NNCI-2025233).

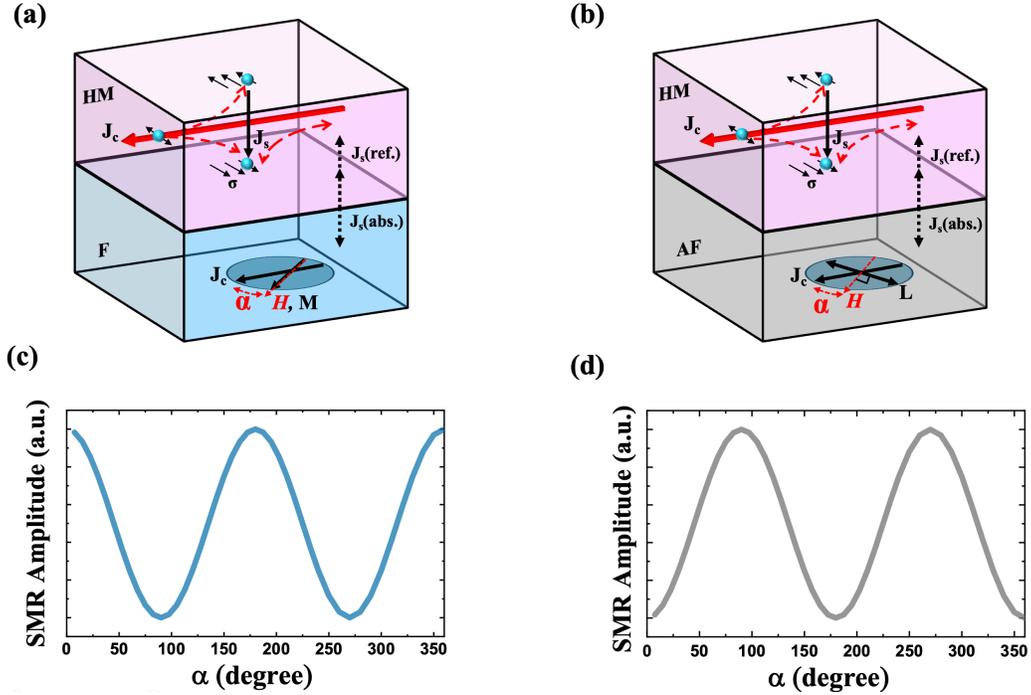

**Fig. 1.** Schematic illustrations of the conventional pictures of spin Hall magnetoresistance (SMR). (a) Illustration of SMR for a heavy metal/ferromagnetic insulator (FMI) bilayer. (b) Illustration for a heavy metal/antiferromagnetic insulator (AFMI) bilayer. Here, $J_c$ is the charge current, $J_s$ is the spin current and α is the relative angle between applied current and the magnetic field directions. (c) For the HM/FMI case the SMR amplitude is maximal when the magnetic-field angle α = 0°, which aligns to magnetization perpendicular to the spin generated by the spin Hall effect. (d) For the HM/AFMI case the conventional picture is that SMR is maximal for α = 90° which favors alignment of the Néel vector perpendicular to the spin generated by the spin Hall effect.

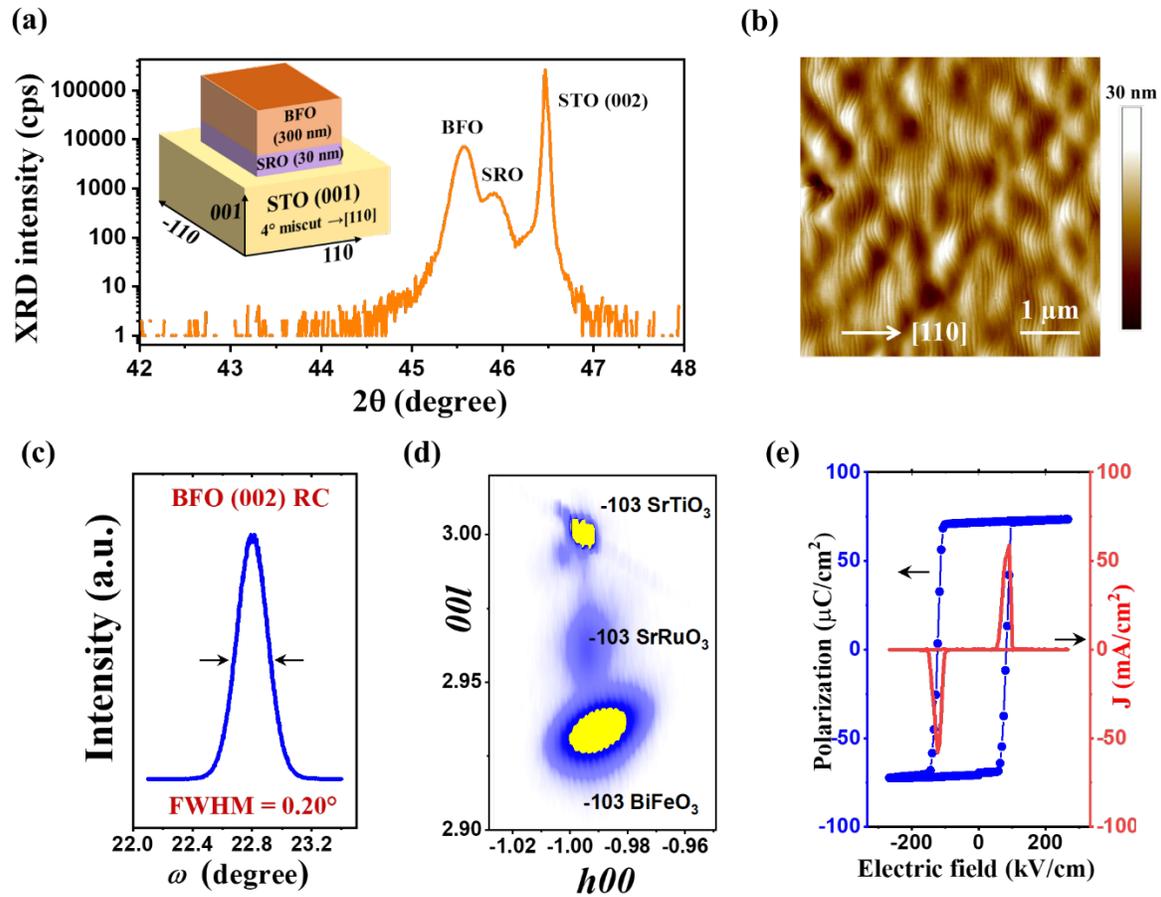

**Fig. 2.** Characterization of the bismuth ferrite (BFO) films. (a) Room-temperature X-ray diffraction spectra around the out-of-plane (002) peak. The inset to (a) shows a schematic representation of a BFO thin film grown on a SrTiO$_3$ (001) substrate with a 4°-miscut toward [110], buffered by a 30 nm thin SrRuO$_3$ layer. (b) Atomic-force-microscopy image showing the surface morphology of the BFO film. (c) Rocking curve around the (002) BFO peak. (d) Reciprocal space map (RSM) around the ($\bar{1}$03) peak showing that the BFO contains only a single ferroelastic domain. (d) Ferroelectric P-E loop taken at a frequency of 10 kHz.

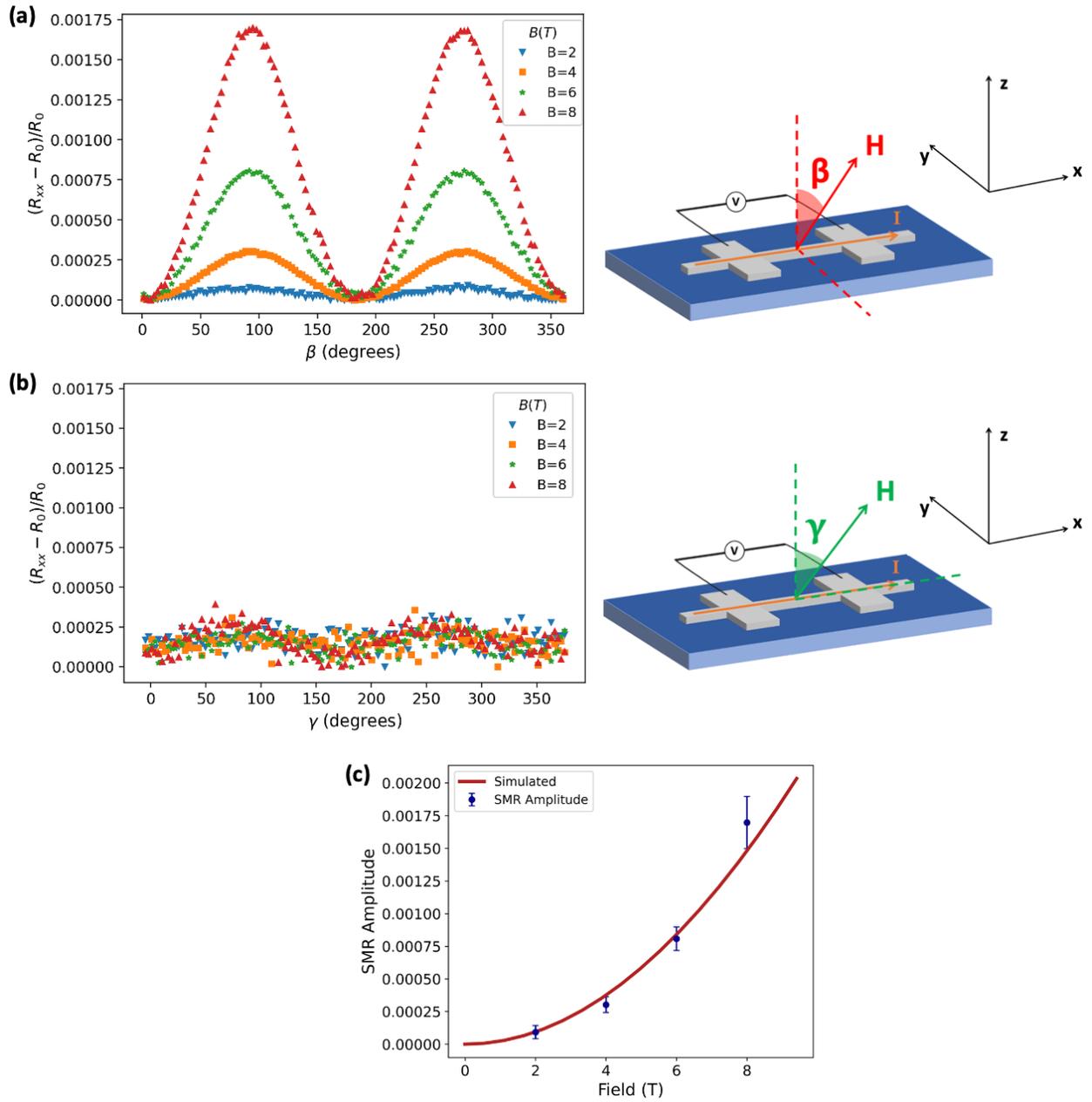

**Fig. 3.** Magnetoresistance for magnetic fields swept away from the sample plane. (a) Longitudinal angle-dependent magnetoresistance measured in a BFO (100 nm)/Pt (10 nm) bilayer as a function of magnetic-field angle in the yz plane ($\beta$ scan) at room temperature. (b)

Longitudinal angle-dependent magnetoresistance measured in the same sample as a function of angle in the xz plane ($\gamma$ scan). (c) Measured SMR amplitude (blue dots) as a function of magnetic field applied in yz plane ($\beta$ scan) plotted against the simulated field dependence for the spin-cycloid model (red curve). (See supplementary material Section S3 for discussion of the simulation.)

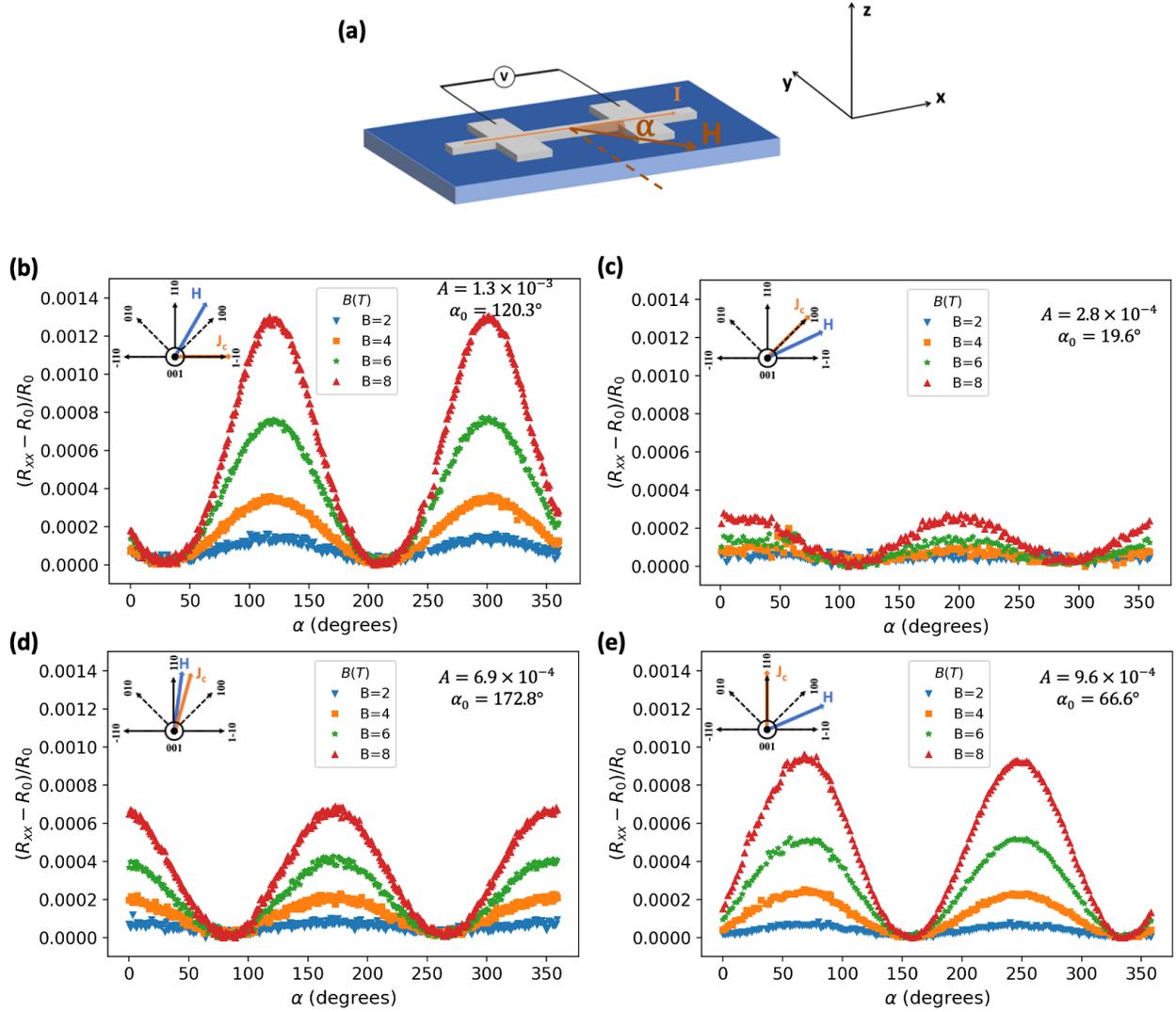

**Fig. 4.** Magnetoresistance for magnetic field swept within the sample plane. (a) Illustration of sample geometry and definition of the in-plane magnetic field angle α. (b)-(e) Longitudinal angle-dependent magnetoresistance measured in four BFO (100 nm)/Pt (10 nm) bilayer samples on the same wafer as a function of magnetic-field angle in the xy plane (α scan). The insets show the direction of current (labeled as $J_C$) for each device and the orientation of magnetic field where the SMR amplitude is maximized, $α_0$ (labeled as $H$). The value indicated for $A$ is the amplitude of the magnetoresistance at 8 T.

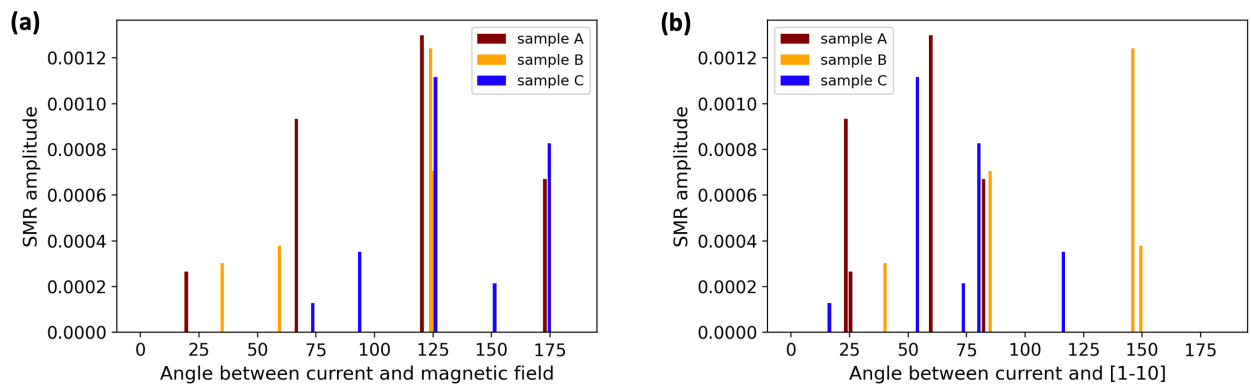

**Fig. 5.** Amplitude of the spin Hall magnetoresistance for different samples. (a) The positions of the lines indicate the magnetic field angles $\alpha_0$ relative to the current direction which give the maximum value of SMR for different devices. The lengths of the lines indicate the corresponding SMR amplitudes, $A$. (b) Same data, but here the positions of the lines indicate the magnetic field angles for maximum SMR relative to the $[1\bar{1}0]$ crystal axis instead of the current direction.

# Unusual dependence on the angle of magnetic field for the spin Hall magnetoresistance of monodomain epitaxial BiFeO$_3$ thin films

*Yongjian Tang[1], Pratap Pal[2], Matthew Roddy[1], Jon Schad[2], Ruofan Li[1], Joongwon Lee[3], Farhan Rana[3], Tianxiang Nan[4], Chang-Beom Eom[2], and Daniel C. Ralph[1,5]\**

[1] Department of Physics, Cornell University, Ithaca, New York 14853, USA.
[2] Department of Materials Science and Engineering, University of Wisconsin-Madison, Madison, Wisconsin 53706, USA.
[3] School of Electrical and Computer Engineering, Cornell University, Ithaca, New York 14853, USA.
[4] School of Integrated Circuits and Beijing National Research Center for Information Science and Technology (BNRist), Tsinghua University, Beijing 100084, China
[5] Kavli Institute at Cornell, Ithaca, New York 14853, USA

*Corresponding Author: Daniel C. Ralph, Email: dcr14@cornell.edu

**Section S1. Additional Data**

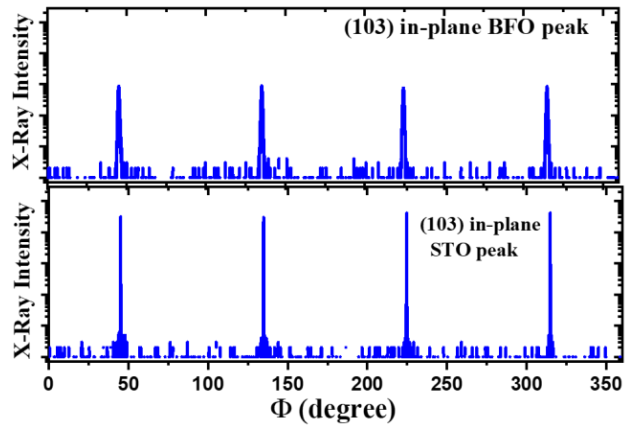

**Figure S1.** X-ray diffraction data: Φ-scans around symmetric (103) peak highlighting epitaxial growth of 300 nm BFO thin-film on STO (001).

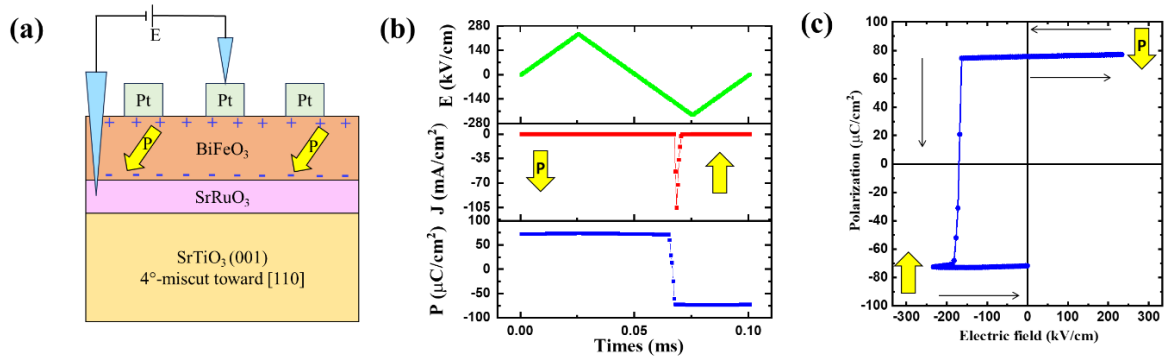

**Figure S2.** (a) Schematic illustration of the sample device used for room-temperature ferroelectric measurements. (b) Variations of the applied electric pulse, the current density and calculated polarization with time. BFO film starts with saturated positive polarization (which by our convention corresponds to a downward-pointing polarization) before switching to up-polarization by a sufficiently large negative electric field. (c) P vs E loop as constructed from above data at 10 kHz frequency indicating a single ferroelectric domain (initially with down polarization) before switching to the opposite polarization.

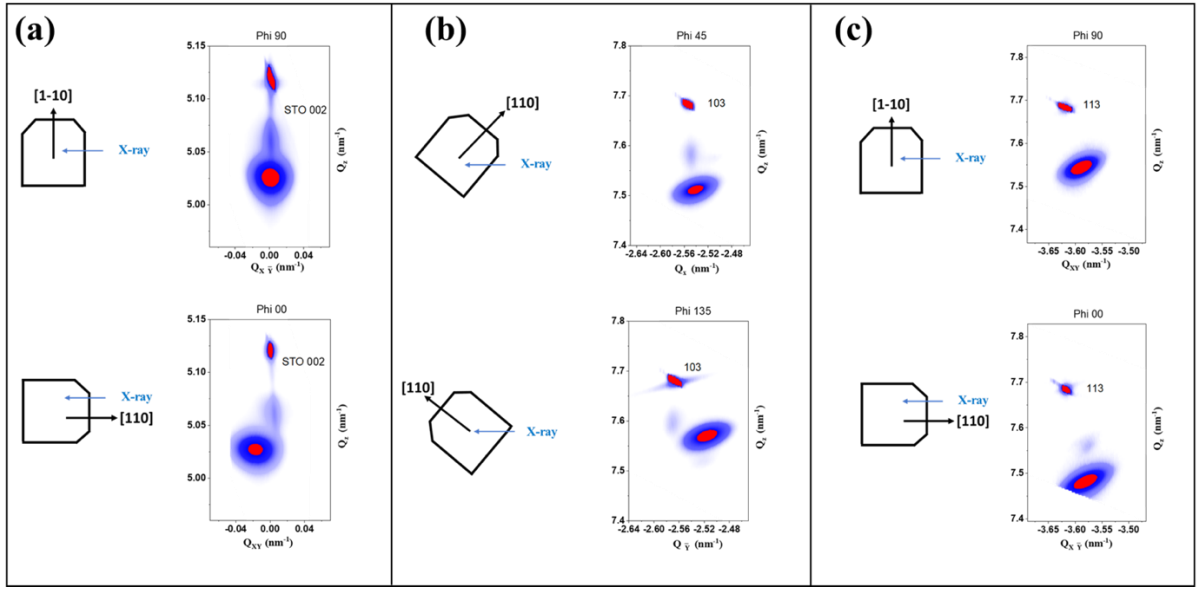

**Figure S3.** Reciprocal space map scans of the $BiFeO_3/SrRuO_3/SrTiO_3$ (001) thin film (a) about the (002) reflection along two in-plane Q-directions, (b) about off-axis (-103) and (c) (-113). All these reflections illustrate the monodomain nature of the BFO films.

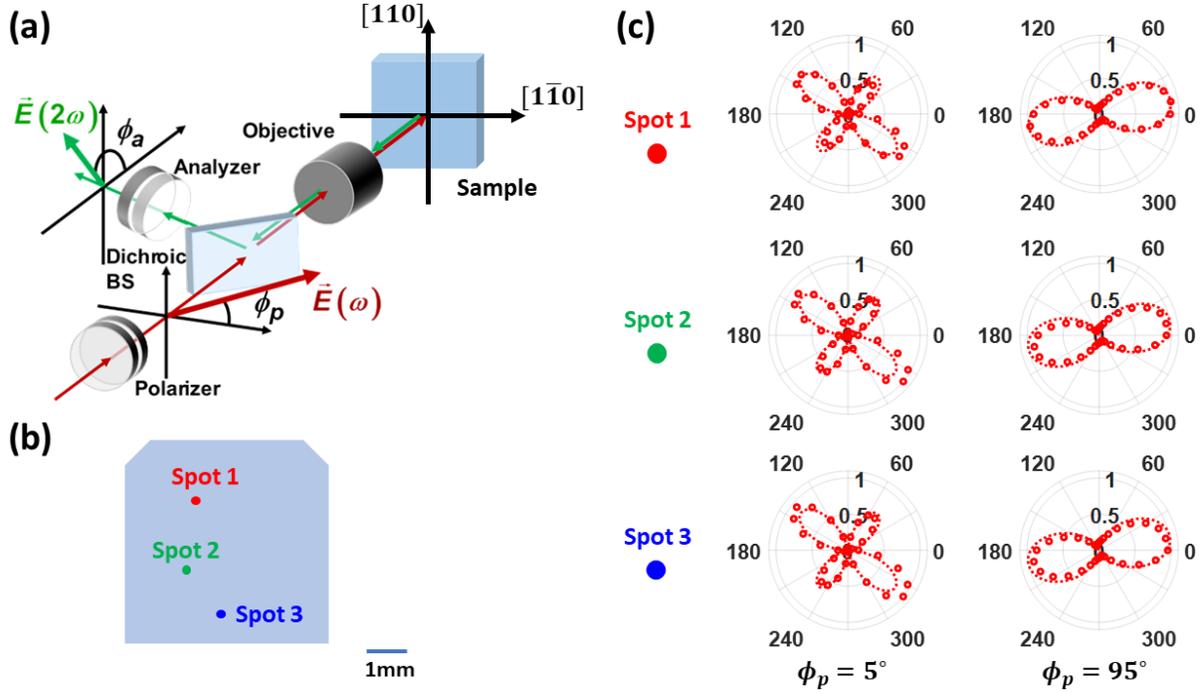

**Figure S4.** (a) Schematic of an optical second harmonic generation (SHG) setup used to check that the BFO layers each contain a single ferroelectric domain. A 1032 nm laser was normally incident to the sample and reflected SHG was measured. (b,c) SHG intensity measured in 3 different spots (shown in (b)) plotted as a function of the analyzer angle while the polarizer angle ($\phi_p$) was fixed at 0° and 90°. These SHG patterns support the conclusion that the sample contains a single [-1-1-1] ferroelectric domain, which is different from the those of the other six domains not along the [-1-1-1] axis. The small intensity difference in the spots measured is attributed to the surface roughness.

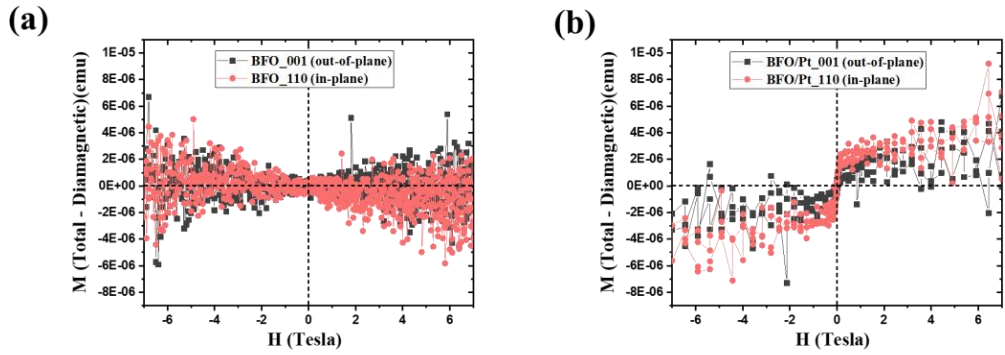

**Figure S5.** Room temperature SQUID measurements (a) before and (b) after Pt deposition in the in-plane and out-of-plane thin-film configurations.

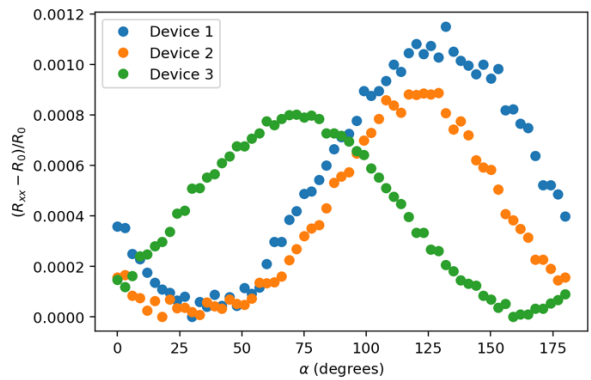

**Figure S6.** SMR measured on three devices separated by 20 μm on the same chip, where electrical current is applied along the same direction. A difference in the phase of the SMR is observed even between samples in close proximity measured together.

**Section S2. Calculation of SMR for an Antiferromagnet with Domains**

Here we will generalize the argument of Fischer *et al*. (*Phys. Rev. B* **2018**, *97*, 014417), who considered the case of an antiferromagnet with 3 equivalent domains in which the in-plane orientations of the Néel vector are equally spaced. We will analyze the case of $N$ equivalent antiferromagnetic domains (with $N \geq 2$), with in-plane orientations for the Néel vector given by the equally-spaced angles $\beta_i = \beta_0 + i\pi/N$, for $i = 0, \ldots, $ N-1, measured from the y axis. By the logic of Fischer *et al*., the difference from equally-weighted fractions for the different domains is given the by field-dependent fractions

$$\Delta \xi_i \propto \cos\left[2\left(\alpha - \left(\beta_i + \frac{\pi}{2}\right)\right)\right] = -\cos[2(\alpha - \beta_i)]$$

where $\alpha$ is the angle of the applied magnetic field, also measured from the y axis. The field-dependent part of the resistance is then

$$\Delta R \propto -\sum_i \cos^2(\beta_i) \cos[2(\alpha - \beta_i)]$$

$$= -\sum_i \cos^2(\beta_i)[\cos 2\alpha \cos 2\beta_i - \sin 2\alpha \sin 2\beta_i]$$

$$= -\cos 2\alpha \sum_i \cos^2(\beta_i)\cos 2\beta_i + \sin 2\alpha \sum_i \cos^2(\beta_i) \sin 2\beta_i$$

$$= -\cos 2\alpha \; \frac{1}{2}\sum_i(1 + \cos 2\beta_i)\cos 2\beta_i + \sin 2\alpha \frac{1}{2}\sum_i(1 + \cos 2\beta_i) \sin 2\beta_i$$

$$= -\cos 2\alpha \; \frac{1}{2}\sum_i(\cos 2\beta_i + \cos^2(2\beta_i)) + \sin 2\alpha \frac{1}{2}\sum_i\left(\sin 2\beta_i + \frac{\sin 4\beta_i}{2}\right).$$

The sums over $\cos 2\beta_i$ and $\sin 2\beta_i$ give zero for any $N$ greater than or equal to 2 as long as the orientation angles are equally spaced. (They are just the total y or x coordinate of a phasor sum that adds to zero.) The sum over $\sin 4\beta_i$ also gives zero for any $N$ greater than or equal to 3, in which case the field-dependent part of the resistance is simply proportional to $-\cos 2\alpha$, independent of the orientations of the easy axes for the domains. (This is the result found by Fischer et al.) However, the case of $N = 2$ is different, because for this case $\sin 4\beta_1 = \sin\left(4\beta_0 + \frac{4\pi}{2}\right) = \sin 4\beta_0$ and the sum over $\sin 4\beta_i$ is not zero. For the case of $N = 2$,

$$\Delta R \propto -\cos 2\alpha \cos^2(2\beta_0) + \sin 2\alpha \frac{1}{2}\sin 4\beta_0$$

$$= -\cos(2\beta_0)\,[\cos 2\alpha \cos(2\beta_0) - \sin 2\alpha \sin(2\beta_0)]$$

$$= -\cos(2\beta_0)\,[\cos[2(\alpha-\beta_0)]].$$

So for the case of $N = 2$ one should not expect the simple result that the SMR resistance has the angular dependence $\Delta R \propto -\cos 2\alpha = \cos[2(\alpha - 90°)]$ independent of the crystal orientation, but instead one should have $\Delta R \propto \cos[2(\alpha-\beta_0)]$ with an explicit dependence on the crystal orientation through the anisotropy angle $\beta_0$.

**Section S3. Simulation of SMR for a Sample Containing a Spin Cycloid**

To determine the dependence of SMR on the magnitude of applied magnetic field for a material containing a spin cycloid, we performed a numerical study of an array of $N = 30$ spins with periodic boundary conditions, with each spin lying initially within the xy plane with an angle from the y-axis ($\theta_i$) governed by the energy function

$$E = \sum_i \left[-H\cos\theta_i - K_{ex}\cos\left(\theta_{i+1} - \theta_i - [\pi + \frac{2\pi}{N}]\right)\right],$$

so that in the absence of an applied magnetic field H the effective magnetic interaction $K_{ex}$ causes the spins take the form of a spin cycloid. We simulated the spin dynamics based on the Landau-Lifshitz-Gilbert equation

$$\frac{d\hat{\boldsymbol{m}}_i}{dt} = -\gamma \hat{\boldsymbol{m}}_i \times \boldsymbol{H}_{\text{eff},i} + \alpha_G \hat{\boldsymbol{m}}_i \times \frac{d\hat{\boldsymbol{m}}_i}{dt}$$

where $\hat{m}_i$ is a unit vector corresponding to spin $i$, $\gamma$ is the gyromagnetic ratio, $\alpha_G$ is the Gilbert damping constant, and

$$\boldsymbol{H}_{\text{eff},i} = \boldsymbol{H} + K_{ex} R\left(\pi + \frac{2\pi}{N}\right)\hat{\boldsymbol{m}}_{i-1} + K_{ex} R\left(\pi - \frac{2\pi}{N}\right)\hat{\boldsymbol{m}}_{i+1} - H_{an} m_{i,z}\hat{\boldsymbol{z}}.$$

Here $R(\phi)$ is a rotation matrix by the angle $\phi$ within the xy plane and the term containing $H_{an}$ models an in-plane anisotropy. The field dependence of the sum of $m_{i,y}^2$ is shown in Figure 3(c) in the main text for an applied magnetic field along the y axis. The result fits closely to a quadratic curve when $H \ll K_{ex}$. Furthermore, if we select a fixed field magnitude within the regime where the dependence is quadratic and rotate the angle $\beta$ of field away from the xy plane, the result for the sum of $m_{i,y}^2$ versus $\beta$ follows a sinusoidal curve with 180° periodicity. The angle dependence corresponds to $\cos 2\beta$, similar to what we observe experimentally.

**Section S4.** Summary of the applied current angles and measured values of α₀ for the samples not already summarized in Fig. 4 of the main text.

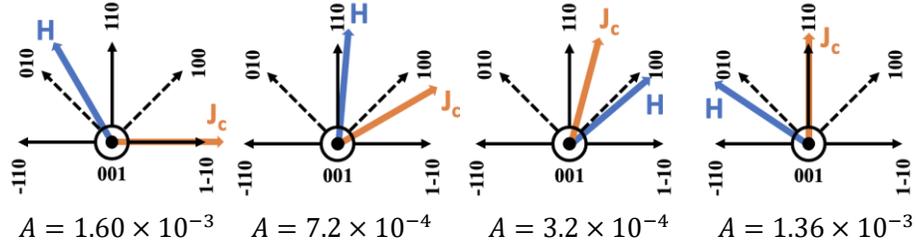

$A = 1.60 \times 10^{-3}$   $A = 7.2 \times 10^{-4}$   $A = 3.2 \times 10^{-4}$   $A = 1.36 \times 10^{-3}$

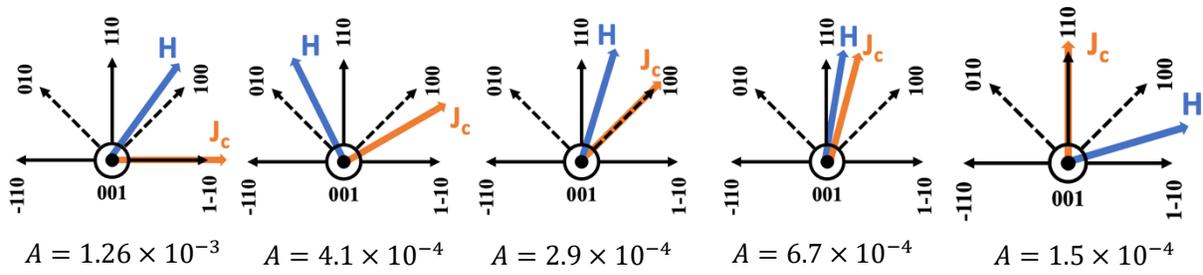

$A = 1.26 \times 10^{-3}$   $A = 4.1 \times 10^{-4}$   $A = 2.9 \times 10^{-4}$   $A = 6.7 \times 10^{-4}$   $A = 1.5 \times 10^{-4}$